\documentclass[12pt]{iopart}
\usepackage{times}
\expandafter\let\csname equation*\endcsname\relax
\expandafter\let\csname endequation*\endcsname\relax
\usepackage{amsmath,amsfonts,mathrsfs}
\usepackage{bbm,dsfont}
\usepackage{graphicx}
\usepackage[shortlabels]{enumitem}
\eqnobysec

\begin{document}
\title{The dual model for an Ising model with nearest and next-nearest neighbors}

\date{\today}
\begin{abstract}
We construct and analyse a dual model to the Ising model with the nearest and next-nearest neighbors
on the rectangular lattice (NNNI model).
The Hamiltonian of the dual model turns out to contain two- and four-spin interactions.
The free fermion approximation suggests that
an increase in the critical temperature of the dual model caused by the four-spin interactions is limited to a finite range.
\end{abstract}
\author{Adam Strycharski, Zbigniew Koza}
\address{Institute of Theoretical Physics, University of Wroc{\l}aw, Poland}
\eads{\mailto{astrycharski@ift.uni.wroc.pl}, \mailto{zkoza@ift.uni.wroc.pl}}
\maketitle

\section{Introduction}
Duality \cite{krammers_wannier,wannier,onsager,syozi} is an important symmetry of
Ising models,
as it implies a relationship between the partition functions of two Ising models with Hamiltonians
 $\mathscr{H}_{1}\left(J\right)$ and $\mathscr{H}_{2}\left(K\right)$
\begin{equation}
 \frac{Z_{1} \left(\mathscr{H}_{1}\left(J\right)\right)}{f_{1}\left(J\right)}
   =
 \frac{Z_{2} \left(\mathscr{H}_{2}\left(K\right)\right)}{f_{2}\left(K\right)},
\end{equation}
where $f_{1}$ and $f_{2}$ are some nonsingular functions of interaction constants $J,K$
appearing in Hamiltonians $\mathscr{H}_{1}$ and $\mathscr{H}_{2}$, respectively.
Similar functional relations between the partition functions of two different  Ising  models
can be also obtained with the help of the  decoration, star-triangle or star-square transformations \cite{syozi}.
Duality combined with those transformations enables one to derive the expressions
for the critical temperature for a wide range of Ising models.
It should be stressed, however, that in the case of duality the total
number of spin interactions must be the same for both Hamiltonians,
$\mathscr{H}_{1}$ and $\mathscr{H}_{2}$,  which allows one to introduce the concept of dual lattice.
For example, the triangular lattice turns out dual to the honeycomb lattice
and the rectangular lattice is dual to itself.
Using the concept of dual lattice, the duality relation can be interpreted
as a relationship between high-temperature (low-temperature) expansion
for the model with Hamiltonian $\mathscr{H}_{1}$  and the low-temperature (high-temperature) expansion
for the model described  by $\mathscr{H}_{2}$.

The concept of duality has been extended by Wegner \cite{wegner} for a wider class of Ising models $M_{dn}$
on d-dimensional lattices characterized by a number $n = 1, 2, \ldots , d$,
where $n = 1$ corresponds to the Ising model with a two-spin interaction.
In this paper we construct and investigate analytically the dual model to the Ising model on a rectangular lattice
with the nearest- and next-nearest-neighbor interactions (NNNI model) \cite{zandvlietOriginal,bittner,lee,kim}.
Although this simple extension of the Onsager model has been widely investigated for decades now,
no analytical expression for the partition function of the model has been found, nor its dual model has been presented.
To this end we use the usual algebraic approach and express the partition function as the trace of a power of the transfer matrix.
In \Sref{sec:1} we employ the classic 2D Ising model on a rectangular lattice to introduce the formalism and some basic formulas necessary to investigate the duality of a more complex model in \Sref{sec:union-jack}. In particular, we show how the self-dual symmetry of this model can be related to the similarity transformation ($U$) of the transfer matrix. We show that the difference between two forms of the transfer matrix
can be related to a change of the direction in which the matrix was constructed and present the explicit form of $U$.
In \Sref{sec:union-jack} we apply the similarity transformation to the transfer matrix of the NNNI model.
We show that there exists  a simple  Ising model for which the transformed matrix is  the transfer matrix.
The Hamiltonian of this model contains a two-spin nearest-neighbor interaction between nodes of the Brick-Wall lattice
(which is topologically equivalent to a honeycomb lattice) and additional four-spin interactions (henceforth this model will be called the (2 + 4) BWI model).
The number of the four-spin interactions is equal to the number of bricks in the lattice.
Since  the two Hamiltonians have the same number of all interactions, the  model can be regarded
as the dual model to the NNNI model.
As a result we obtain the duality relation between the partition functions of these  two models.
Moreover,  we show that  the dual model  becomes  self-dual in the presence of an external magnetic field.
In ~\Sref{sec:tc} the critical temperature of the (2 + 4) BWI model is investigated.
We show that an increase in the critical temperature of the dual model caused by the four-spin interactions
is limited to a finite range. We derive the equations for both the lower and upper limits of this temperature range and find that they depend on the magnitude of the two-spin interactions only.

%%%%%%%%%%%%%%%%%%%%%%%%%%%%%%%%%%%%%%%%%%%%%%%%%%%%%%%%%%%%%%%%%%%%%%%
%%%%%%%%%  SECTION 1  %%%%%%%%%%%%%%
%%%%%%%%%%%%%%%%%%%%%%%%%%%%%%%%%%%%%%%%%%%%%%%%%%%%%%%%%%%%%%%%%%%%%%%

\section{Selfduality of the Ising model on the rectangular lattice \label{sec:1}}
Consider the classical Ising model with the nearest-neighbor interaction constants $J_{1}$ and $J_{2}$ (for simplicity, we assume that $J = J_{1} = J_{2}$)
on a rectangular lattice with  $N$ columns and $2M$ rows.
We assume the cyclic boundary conditions in the direction of the transfer matrix action
and in the perpendicular direction we adjust  the boundary conditions to the transformations
performed on the transfer matrix to get the result in a closed form
(we adopt this convention throughout this paper).
The partition function for this model reads \cite{schultz, onsager, thompson}
\begin{eqnarray}
\label{Z_1}
Z_{N,2M}(\kappa) &=& \operatorname{Tr}V^{N} \nonumber \\
               &=& \left(2 \sinh 2\kappa\right)^{MN} \operatorname{Tr}\left(e^{\kappa A}e^{\tilde{\kappa}C}\right)^{N},
\end{eqnarray}
where $\kappa = \beta J = J/k_{\mathrm{B}} T$,
$\tilde{\kappa} = -\frac{1}{2} \ln \tanh \kappa$,
and $A$ and $C$ are $2^{2M}\times 2^{2M}$ matrices defined by
\begin{eqnarray}
  A &=& \sigma_{1}^{x} + \sum\limits_{j = 1}^{2M - 1}{\sigma_{j}^{x} \sigma_{j+1}^{x}},\\
  C &=& \sum\limits_{j = 1}^{2M}{\sigma_{j}^{z}},
\end{eqnarray}
with
\begin{equation}
  \sigma_{j}^{\alpha} =
   \underbrace{ \mathds{1} \otimes \mathds{1} \otimes \ldots \otimes \mathds{1} }_{j-1}
      \otimes \sigma^{\alpha} \otimes
   \underbrace{\mathds{1} \otimes \ldots \otimes \mathds{1}}_{2M-j},
\end{equation}
where the symbol $\otimes$ denotes the tensor product, $\mathds{1}$ is the $2 \times 2$ unit matrix,
and $\sigma^{\alpha}$ ($\alpha = x, y, z$) represent the Pauli matrices
\begin{equation}\sigma^{x} =
\left(
\begin{array}{c c}
		0 & 1 \\
		1 & 0 \\
		\end{array}
\right),
\sigma^{y} =
\left(
\begin{array}{c c}
		0 & -i \\
		i & 0 \\
		\end{array}
\right),
\sigma^{z} =
\left(
\begin{array}{c c}
		1 & 0 \\
		0 & -1 \\
		\end{array}
\right),
\end{equation}
and $j$ runs through the $2M$ nodes of a lattice column.

It turns out that $A$ and  $C$ a related to each other by a similarity transformation $U$,
\begin{equation}
\label{eq:similarity}
   UAU^{-1} = C,  \qquad
   UCU^{-1} = A.
\end{equation}
Relations of this type were found to be  essential in proving the duality
or self-duality of many Ising models, including the self-duality of
the 1D Ising model with an external transverse magnetic field \cite{fradkin_susskind,kogut,turban}.

Applying \Eref{eq:similarity} under the trace operator in \Eref{Z_1}, one obtains
\begin{eqnarray}
\label{Eq:up_form_z}
 Z_{N,2M}(\kappa) &=&
   (2 \sinh 2\kappa)^{MN} \operatorname{Tr} \left( U \left(e^{\kappa A} e ^{\tilde{\kappa} C}\right)^{N} U^{-1} \right) \nonumber \\
  &=&  (2 \sinh 2\kappa)^{MN} \operatorname{Tr}  \left(e^{\kappa U A U^{-1}} e^{\tilde{\kappa} U C U^{-1}}\right)^{N} \nonumber \\
  &=& (2 \sinh 2\kappa)^{MN} \operatorname{Tr}  \left(e^{\kappa C} e^{\tilde{\kappa} A}\right)^{N}.
\end{eqnarray}
Using the cyclicity of the trace operator, \Eref{Eq:up_form_z} can be rewritten in a well-known form
\begin{eqnarray}
 \label{eq:prof-2D}
  Z_{N,2M}(\kappa) &=&  (2 \sinh 2\kappa)^{MN} \operatorname{Tr}  \left(e^{\tilde{\kappa} A} e^{\kappa C}\right)^{N} \nonumber \\
  & = &  \left(\frac{2 \sinh 2 \kappa}{2 \sinh 2 \tilde{\kappa}}\right)^{MN} (2 \sinh 2 \tilde{\kappa})^{MN} \operatorname{Tr} \left(e^{\tilde{\kappa} A} e^{\kappa C}\right)^{N}  \nonumber\\
  & = &  \left(\frac{2 \sinh 2 \kappa}{2 \sinh 2 \tilde{\kappa}}\right)^{MN} Z_{N,2M}(\tilde{\kappa})  \nonumber\\
  & = &  (\sinh 2 \kappa)^{2MN} Z_{N,2M}(\tilde{\kappa}),
\end{eqnarray}
which proves the self-duality of the discussed model.

\Eref{eq:prof-2D} was usually derived by constructing the transfer matrix
along different directions of the lattice, without deriving the explicit form of $U$,
which however will be necessary in our study further below.
To obtain it, one can assume that it can be expressed as a product of two $2^{2M} \times 2^{2M}$ matrices
\begin{equation}
 \label{eq:SR}
   U = SR,
\end{equation}
where $R$ is a permutation matrix,
whereas $S$ exchanges Pauli matrices  $\sigma^{x}_{i}$ with $\sigma^{z}_{i}$ and changes the signs
of the interaction coefficients appropriately.
A particularly simple form of $R$ reads
\begin{equation}
 \label{eq:R_1}
   R = \prod\limits_{j = 1}^{2M - 1}{R_{j,j + 1}},
\end{equation}
where
\begin{equation}
  R_{j, j + 1} =
  \underbrace{\mathds{1} \otimes \mathds{1} \otimes \ldots  \otimes \mathds{1}}_{j-1} \otimes r \otimes
  \underbrace{\mathds{1} \otimes \ldots \otimes \mathds{1}}_{2M-j-1}
\end{equation}
and
\begin{equation}
	r = \left(
		\begin{array}{c c c c}
		0 & 1 & 0 & 0 \\
		1 & 0 & 0 & 0 \\
		0 & 0 & 1 & 0 \\
		0 & 0 & 0 & 1	
		\end{array}
	\right ).
\end{equation}
It follows that $r^{-1} = r$, $[R_{j,j+1},R_{j+1,j+2}] \neq 0$, and
\begin{equation}
  \label{eq:R-1_1}
    R^{-1} = \prod\limits_{l = 1}^{2M - 1}{R_{2M - l,2M - (l - 1)}}.
\end{equation}
The following relations
\begin{eqnarray}
R_{l, l + 1} \sigma_{l}^{x} \sigma_{l+1}^{x} R_{l, l + 1}^{-1} &=& \sigma_{l + 1}^{x} \\ \nonumber
R_{l, l + 1} \sigma_{l}^{x} R_{l, l + 1}^{-1} &=& \sigma_{l}^{x} \\ \nonumber
R_{l, l + 1} \sigma_{l+1}^{x} R_{l, l + 1}^{-1} &=& \sigma_{l}^{x} \sigma_{l+1}^{x} \\ \nonumber
R_{l, l + 1} \sigma_{l}^{z} \sigma_{l+1}^{z} R_{l, l + 1}^{-1} &=& - \sigma_{l}^{z} \\ \nonumber
R_{l, l + 1} \sigma_{l}^{z} R_{l, l + 1}^{-1} &=& -\sigma_{l}^{z} \sigma_{l+1}^{z} \\ \nonumber
R_{l, l + 1} \sigma_{l+1}^{z} R_{l, l + 1}^{-1} &=& \sigma_{l+1}^{z}
\end{eqnarray}
can be used to show how $A$ and $C$ transform under  $R$,
\begin{eqnarray}
\label{eq:R_2}
 RAR^{-1} &=&
    \sum\limits_{j = 2}^{2M} \sigma_{j}^{x}, \nonumber\\
 RCR^{-1} &=&
    \sum\limits_{j = 1}^{2M - 1} \sigma_{l}^{z}\sigma_{j + 1}^{z} + \sigma_{2M}^{z}.
\end{eqnarray}

The cyclic boundary conditions in  $A$ are mathematically troublesome.
However, since the thermodynamics of the system does not depend on the boundary conditions in the thermodynamic limit,
they can be neglected \cite{fradkin_susskind,kogut}.
As for $S$, it can be  expressed as
\begin{equation}
   S = \mathscr{P}\prod\limits_{j = 1}^{2M} s_{j} ,
\end{equation}
where
\begin{equation}
\label{eq:sj}
  s_{j} = \underbrace{\mathds{1} \otimes \mathds{1} \otimes \ldots \mathds{1}}_{j-1}
     \otimes s \otimes
  \underbrace{\mathds{1} \otimes \ldots \otimes \mathds{1}}_{2M-j},
\end{equation}
with
\begin{equation}
\label{eq:s}
s = \frac{1}{\sqrt{2}}\left(
\begin{array}{c c}
		1 & 1 \\
		1 & -1 \\
		\end{array}
\right),
\end{equation}
whereas $\mathscr{P}$ is a transformation exchanging matrices  $\sigma^{\alpha}_{i}$ with $\sigma^{\alpha}_{2M - i + 1}$,
\begin{equation}
\mathscr{P} = \prod_{i = 1}^{M}{P_{i,2M - i + 1}},
\end{equation}
where
\begin{equation}
P_{i,j} = \frac{1}{2}\left(1 + \sigma_{i}^{x}\sigma_{j}^{x} + \sigma_{i}^{y}\sigma_{j}^{y} + \sigma_{i}^{z}\sigma_{j}^{z}\right)
\end{equation}
and
\begin{equation}
P_{i,j} \sigma_{i}^{\alpha} P_{i,j}^{-1} = \sigma_{j}^{\alpha}, \mbox{ } \alpha = x,y,z \mbox{ and } i,j = 1,2,\ldots, 2M.
\end{equation}
On applying the full transformation $U = SR$  to $A$ and $C$, one arrives at \Eref{eq:similarity}.

%%%%%%%%%%%%%%%%%%%%%%%%%%%%%%%%%%%%%%%%%%%%%%%%%%%%%%%%%%%%%%%%%%%%%%%
%%%%%%%%%  SECTION 2  %%%%%%%%%%%%%%
%%%%%%%%%%%%%%%%%%%%%%%%%%%%%%%%%%%%%%%%%%%%%%%%%%%%%%%%%%%%%%%%%%%%%%%

\section{The dual model to the NNNI Model \label{sec:union-jack}}
\begin{figure}
\centering
  \includegraphics[width=0.9\columnwidth]{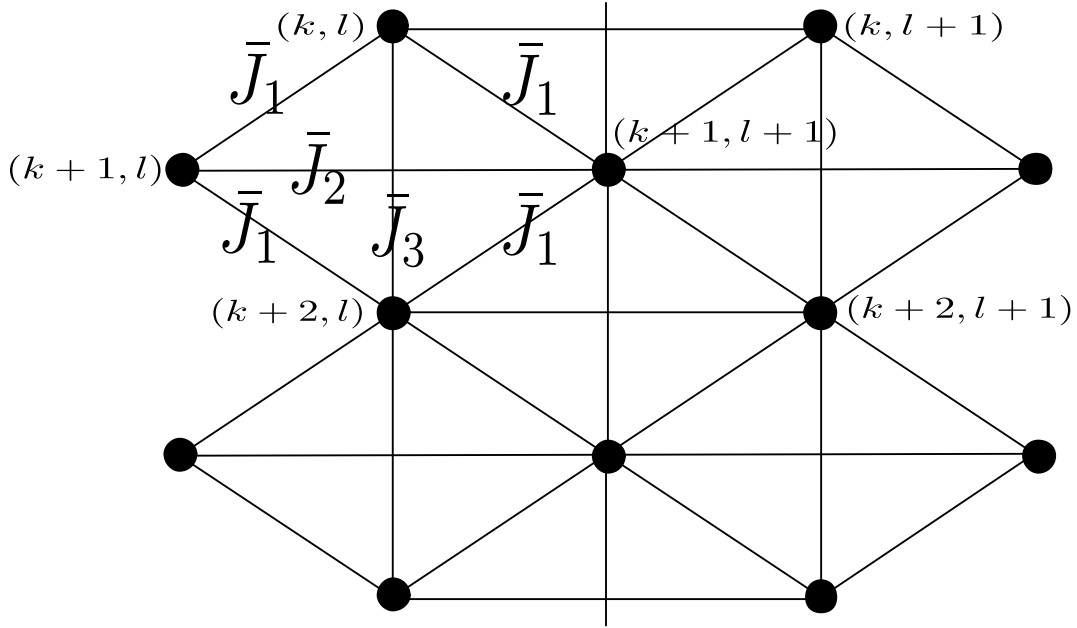}
  \caption{The Ising model with the nearest and next-nearest neighbor interactions
  on a rhomboidal lattice (the NNNI model).
  $\bar{J_{1}}$, $\bar{J_{2}}$, and  $\bar{J_{3}}$ are the interaction constants for the diagonal,
  horizontal and vertical directions, respectively.  }
  \label{fig:lattice_1_2_3}
\end{figure}

The transfer matrix for the Ising model with the nearest  and next-nearearest neighbor
interactions was found in 1956 by Temperley \cite{temperley_book},
who  constructed it along the diagonal of the basic rectangular lattice
defined by the nearest-neighbor interactions.
Although this matrix is not of the lowest possible rank,
its form is particularly simple. We will use this matrix for the rhomboidal
lattice presented in \Fref{fig:lattice_1_2_3}.
While in most papers devoted to the NNNI model only two interaction constant are used \cite{bittner, lee},
$\bar{J_{1}}$, $\bar{J_{2}} = \bar{J_{3}}$, here we consider the general case
with three arbitrary constants controlling  the ferromagnetic interactions
$\bar{J_{1}}, \bar{J_{2}}, \bar{J_{3}}$ ($\bar{J_{i}} < 0$).

\begin{figure}
  \centering
  \includegraphics[width=1.0\columnwidth]{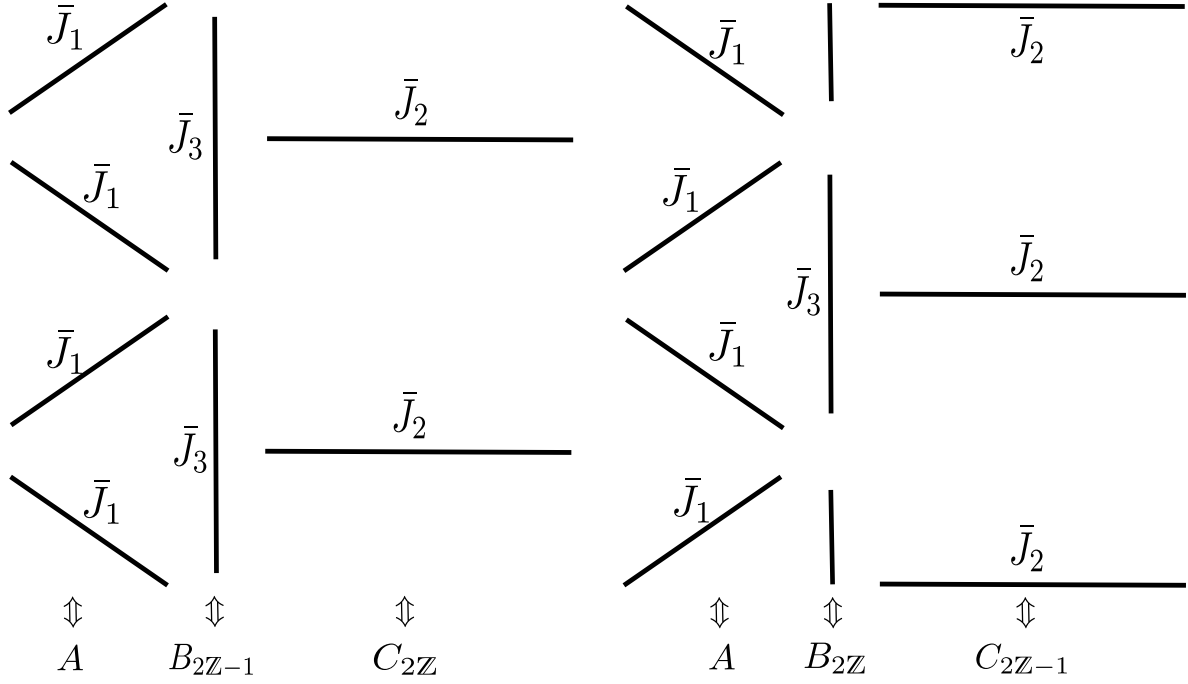}
  \caption{Correspondence between the two-spin interactions and the matrix operators.}
  \label{fig:lattice_1_2_3_d}
\end{figure}

The transfer matrix and the partition function for this model on the rhomboidal  lattice
consisting of $2M \times N$
nodes satisfy
\begin{equation}
\frac{Z_{N,2M}}{\left(2 \sinh 2 L_{2}\right)^{NM}} =
 \operatorname{Tr} \left(e^{L_{1}A} e^{L_{3} B_{2\mathds{Z}-1}}
    e^{\tilde{L_{2}} C_{2\mathds{Z}}} e^{L_{1}A} e^{L_{3} B_{2\mathds{Z}}} e^{\tilde{L_{2}} C_{2\mathds{Z}-1}}\right)^{N},
\end{equation}
where $L_{i} = \beta \bar{J_{i}}$ $(i=1,2,3)$,
$\tilde{L}_{2} = -\frac{1}{2} \ln \tanh L_{2}$,
 $A$, $B_{2\mathds{Z}}$, $B_{2\mathds{Z}-11}$, $C_{2\mathds{Z}}$, and $C_{2\mathds{Z}-1}$
are matrices of size $2^{2M} \times 2^{2M}$, defined as follows,
\begin{eqnarray}
 A                 &=& \sigma_{1}^{x} + \sum\limits_{k = 1}^{2M - 1} \sigma_{k}^{x} \sigma_{k+1}^{x},  \\
 B_{2\mathds{Z}}   &=& \sigma_{2}^{x} + \sum\limits_{k = 1}^{M - 1} \sigma_{2k}^{x} \sigma_{2k+2}^{x}, \\
 B_{2\mathds{Z}-1} &=& \sum\limits_{k = 1}^{M - 1} \sigma_{2k - 1}^{x} \sigma_{2k+1}^{x} + \sigma_{1}^{x} \sigma_{2M - 1}^{x} \sigma_{2M}^{x}, \\
 C_{2\mathds{Z}}   &=& \sum\limits_{k = 1}^{M} \sigma_{2k}^{z}, \\
 C_{2\mathds{Z}-1} &=& \sum\limits_{k = 1}^{M} \sigma_{2k - 1}^{z},
\end{eqnarray}
with $2\mathds{Z}$ denoting the set of even numbers.
The correspondence between these matrices and the interaction constants are shown in  \Fref{fig:lattice_1_2_3_d}.
Similar transfer matrices for the NNNI model were already used in \cite{oitmaa},
however, without giving their explicit forms.

On applying the similarity transformation $U$
to the matrices of the NNNI model we arrive at
\begin{eqnarray}
UAU^{-1}                 &=& \sum\limits_{k = 1}^{2M} \sigma_{k}^{z} = C,  \\
UB_{2\mathds{Z}-1}U^{-1} &=& \sum\limits_{k = 2}^{M} \sigma_{2k - 2}^{z} \sigma_{2k-1}^{z} + \sigma_{1}^{z} \sigma_{2M}^{z} = D_{2\mathds{Z}}, \\
UC_{2\mathds{Z}}U^{-1}   &=& \sigma_{1}^{x} + \sum\limits_{k = 1}^{M - 1} \sigma_{2k}^{x} \sigma_{2k+1}^{x} = A_{2\mathds{Z}}, \\
UB_{2\mathds{Z}}U^{-1}   &=& \sum\limits_{k = 1}^{M} \sigma_{2k-1}^{z} \sigma_{2k}^{z} = D_{2\mathds{Z}-1}, \\
UC_{2\mathds{Z}-1}U^{-1} &=& \sum\limits_{k = 1}^{M} \sigma_{2k - 1}^{x} \sigma_{2k}^{x} = A_{2\mathds{Z}-1},
\end{eqnarray}
which also defines $A_{2\mathds{Z}}$,  $A_{2\mathds{Z}-1}$, $D_{2\mathds{Z}}$, and $D_{2\mathds{Z}-1}$.
These relations lead to
\begin{equation}
  \label{eq:duality-UJ}
  \frac{Z_{N,2M}}{\left(2 \sinh 2 L_{2}\right)^{NM}} = \operatorname{Tr} \tilde{V}^{N},
\end{equation}
where
\begin{equation}
	\label{eq:tildeV}
  \tilde{V} =
    e^{L_{1}C}e^{L_{3}D_{2\mathds{Z}}}e^{\tilde{L_{2}}A_{2\mathds{Z}}}
    e^{L_{1}C}e^{L_{3}D_{2\mathds{Z}-1}}e^{\tilde{L_{2}}A_{2\mathds{Z}-1}}.
\end{equation}

\begin{figure}
	\centering
		\includegraphics[width=0.75\columnwidth]{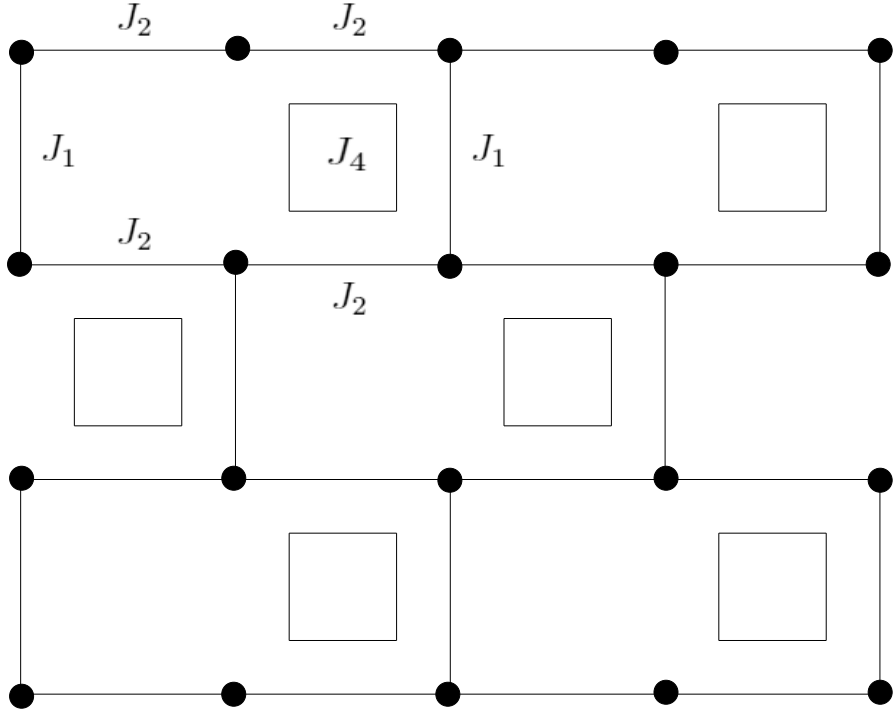}
	\caption{The (2 + 4) BWI Model.}
	\label{fig:square_j4_j1_j2_b}
\end{figure}

Equation (\ref{eq:duality-UJ}) ensures that if there exists an Ising model whose transfer matrix
is equal to $\tilde{V}$, this model will be dual to the NNNI model.
This is a nontrivial requirement, as while the  transfer matrix of any Ising model can be expressed as
a product of exponential matrices,
the product of  exponential matrices is hardly ever the transfer matrix of an Ising model.
In the case considered here, however, such a model exists and is described by the Hamiltonian
\begin{equation}
\mathscr{H} = - \sum\limits_{<i,j>} J_{ij}\sigma_{i}\sigma_{j} - J_{4} \sum\limits_{<i,j,k,l>}
\sigma_{i}\sigma_{j}\sigma_{k}\sigma_{l}
\label{eq:H_J_J4}
\end{equation}
defined on the Brick-Wall lattice \cite{temperley_chapter}, which is topologically equivalent
to the honeycomb lattice.
As explained in \Fref{fig:square_j4_j1_j2_b},
$J_4$ is the four-spin interaction constant for the four spins occupying the right-hand side of a brick,
and the values of the nearest-neighbor interaction constants $J_{ij}$ are non-zero only if
$i$ and $j$ are on the same side of a brick, in which case they are equal $J_1$ and $J_2$
for interactions along the vertical and horizontal direction, respectively.
All interaction constants are ferromagnetic in nature ($J_1, J_2, J_4 < 0$).

The dual lattice is composed of $2M$ horizontal chains, each containing $2N$ nodes.
The partition function of the model satisfies
\begin{eqnarray}
\label{eq:Z_K1_K2_K4}
& &\frac{Z_{2N,2M} \left(K_{2},K_{4},K_{1}\right)}{2^{2MN}\left[\sinh^{2}
    2 K_{2} \left(e^{4K_{4}}\cosh^{2} 2 K_{2} - 1\right)\right]^{MN/2}} = \nonumber\\
& & = \operatorname{Tr} \left(e^{\tilde{K_{2}}C}e^{\tilde{K_{4}}D_{2\mathds{Z}}}e^{K_{1}A_{2\mathds{Z}}}e^{\tilde{K_{2}}C}e^{\tilde{K_{4}}D_{2\mathds{Z}-1}}e^{K_{1}A_{2\mathds{Z}-1}}\right)^{N},
\end{eqnarray}
where $K_{i} = \beta J_{i}$, $(i = 1,2,4)$ and
\begin{eqnarray}
 \label{eq:tilde_K2}
  \tilde{K_{2}} &=& - \frac{1}{4} \ln \frac{\cosh 2K_{2} - e^{-2K_{4}}}{\cosh 2 K_{2} + e^{-2K_{4}}}, \\
 \label{eq:tilde_K4}
  \tilde{K_{4}} &=& - \frac{1}{4} \ln \frac{\cosh^{2} 2K_{2} - 1}{\cosh^{2} 2 K_{2} - e^{-4K_{4}}}.
\end{eqnarray}

The transfer matrix  in \Eref{eq:Z_K1_K2_K4} acts on the vertical groups of nodes and advances
the partition function in the horizontal direction \cite{baxter}.
Its explicit form can be obtained from Onsager's classical transfer matrix for the rectangular lattice
\cite{onsager}, see \Eref{Z_1}. First, we add the four-spin interactions that modify
only two horizontal two-spin bonds lying one above the other.
Next, half of the vertical two-spin interactions connecting the horizontal
chains has to be removed.
While adding the four-spin interactions to the transfer matrix, we  can use a relationship
\begin{equation}
\Delta^{1/4} e^{\tilde{K_{2}}\left(\sigma_{k}^{z} + \sigma_{k+1}^{z}\right)}e^{\tilde{K_{4}}\sigma_{k}^{z}\sigma_{k+1}^{z}}
= \cosh K_{4} c_{k} c_{k+1} + \sinh K_{4} \sigma_{k}^{x} \sigma_{k+1}^{x} c_{k}c_{k+1} \sigma_{k}^{x} \sigma_{k+1}^{x},
\end{equation}
where
$\Delta = \frac{1}{16} \sinh^{2} 2K_{2} \left(e^{4K_{4}} \cosh^{2} 2K_{2} - 1\right)$ and
\begin{equation}
c_{k} = \underbrace{\mathds{1} \otimes \mathds{1} \otimes \ldots \otimes \mathds{1}}_{\mbox{k - 1}}
  \otimes c \otimes
        \underbrace{\mathds{1} \otimes \ldots \otimes \mathds{1}}_{\mbox{2M - k}},
\end{equation}
\begin{equation}
c =
\left(
\begin{array}{c c}
		\cosh K_{2} & 0 \\
		0 & \sinh K_{2} \\
		\end{array}
\right),
\end{equation}
and $k=1,\ldots,\mbox{2M}$.
Finally, we find the duality relationship between the NNNI model, \Fref{fig:lattice_1_2_3},
and the (2 + 4) BWI model, \Fref{fig:square_j4_j1_j2_b},
\begin{equation}
 \label{eq:duality-UJ-BM}
\frac{Z_{N,2M}\left(L_{1},L_{3},L_{2}\right)}{\left(2 \sinh 2 L_{2}\right)^{MN}} =
\frac{Z_{2N,2M} \left(K_{2}, K_{4}, K_{1}\right)}{2^{2MN} \left[\sinh^{2} 2K_{2}
 \left(e^{4K_{4}}\cosh^{2} 2 K_{2} - 1\right)\right]^{MN/2}},
\end{equation}
where the relations between the interaction constants for both model are given by
\begin{equation}
 \label{LLL-KKK}
  L_{1}         = \tilde{K_{2}}, \qquad
  L_{3}         = \tilde{K_{4}}, \qquad
  \tilde{L_{2}} = K_{1}.
\end{equation}

%%%%%%%%%%%%%%%%%%%%%%%%%%%%%%%%%%%%%%%%%%%%%%%%%%%%%%%%%%%%%%%%%%%%%%%
%%%%%%%%%  SECTION 3  %%%%%%%%%%%%%%
%%%%%%%%%%%%%%%%%%%%%%%%%%%%%%%%%%%%%%%%%%%%%%%%%%%%%%%%%%%%%%%%%%%%%%%

\section{The critical temperature for the (2 + 4) BWI model \label{sec:tc}}
While the Ising model on the Brick-Wall lattice can be solved exactly \cite{wannier_brickwall}, addition of the four-spin interaction
makes it intractable analytically.
However, it turns out that  it is still possible to draw some important conclusions about the way the  critical temperature
of the Ising model defined by Hamiltonian (\ref{eq:H_J_J4})
depends on the strength of the four-spin interactions.

%%%%%%%%%%%%%%%%%%%%%%%%%%%%%%%%%%%%%%%%%%%%%%%%%%%%%%%%%%%%%%%%%%%%%%%%%%%%%%%%%%%%%%%%%%%%%%%%%%%%%%%%%%%%%%%%%%%%%%%%%%%%%%%
In the Ising model with nearest  and next-nearest  neighbor interactions (NNNI model, \Fref{fig:lattice_1_2_3})
there are two important limiting cases: $\bar{J_{3}} = 0$ and $\bar{J_{1}} = 0$.
In the former case
the NNNI model reduces
to the Ising model with nearest-neighbor interactions on the  triangular lattice with the interaction constants
$(\bar{J_{1}}, \bar{J_{2}})$.
Since in this case $L_{3} = \beta \bar{J_{3}} = 0$,
matrices $e^{L_{3}D_{2\mathds{Z}}}$ and $e^{L_{3}D_{2\mathds{Z} - 1}}$
appearing in \Eref{eq:tildeV} become the  identity matrices and
the critical point of this model can be determined exactly.
It is worth noticing  that the duality transformation converts the transfer matrix of this model
into the transfer matrix of the Ising model with two nearest-neighbor interactions
on a honey comb lattice equivalent to the Brick-Wall lattice.
This explains why the dual model to NNNI model is defined on the  Brick-Wall lattice.
In the second case ($\bar{J_{1}} = 0$),
the NNNI model splits into two identical independent self-dual Ising models
with nearest-neighbor interactions on the rectangular lattice.
Similarly to the previous case, matrices $e^{L_{1}A}$ in \Eref{eq:tildeV}
become the identity matrices and the critical point for this model is exactly determinable.

\Eref{LLL-KKK},
which describes relationships between  the interactions of the NNNI  and (2+4) BWI models, reduces to
\begin{equation}
\label{eq:L3K4}
0 = L_{3} = \tilde{K_{4}} = - \frac{1}{4} \ln \frac{\cosh^{2} 2K_{2} - 1}{\cosh^{2} 2 K_{2} - e^{-4K_{4}}}
\end{equation}
for  $\bar{J_{3}} = 0$  and to
\begin{equation}
\label{eq:L1K2}
0 = L_{1} = \tilde{K_{2}} = - \frac{1}{4} \ln \frac{\cosh 2 K_{2} - e^{-2K_{4}}}{\cosh 2 K_{2} + e^{-2K_{4}}}
\end{equation}
for  $\bar{J_{1}} = 0$.
\Eref{eq:L3K4} is equivalent to
$K_{4} = 0$, whereas \Eref{eq:L1K2} can be satisfied only in the limit of $K_{4} \rightarrow \infty$.
The critical temperatures satisfy \cite{syozi}
\begin{eqnarray}
\label{eq:e_cosh_tanh}
e^{-4 \frac{J_{4}}{kT_{C}}} &=& \cosh^{2} \left(\frac{2 J_{2}}{kT_{C}}\right) \frac{\tanh^{2}
  \left(\frac{J_{1}}{kT_{C}}\right)}{\tanh^{2} \left(\frac{J_{1}}{kT_{C}}\right) + 1},          \\
\label{eq:e_Cosh_e_sinh}
e^{-4 \frac{J_{4}}{kT_{C}}} &=&
  \cosh^{2} \left(\frac{2 J_{2}}{kT_{C}}\right) - e^{4 \frac{J_{1}}{kT_{C}}} \sinh^{2} \left(\frac{2 J_{2}}{kT_{C}}\right),
\end{eqnarray}
for $\bar{J_{3}} = 0$  and  $\bar{J_{1}} = 0$, respectively.

%%%%%%%%%%%%%%%%%%%%%%%%%%%%%%%%%%%%%%%%%%%%%%%%%%%%%%%%%%%%%%%%%%%%%%%%%%%%%%%%%%%%%%%%%%%%%%%%%%%%%%%%%%%%%%%%%%%%%%%%%%%%%%%%%%%%%%%%%%%

\begin{figure}
	\centering
		\includegraphics[width=0.9\columnwidth]{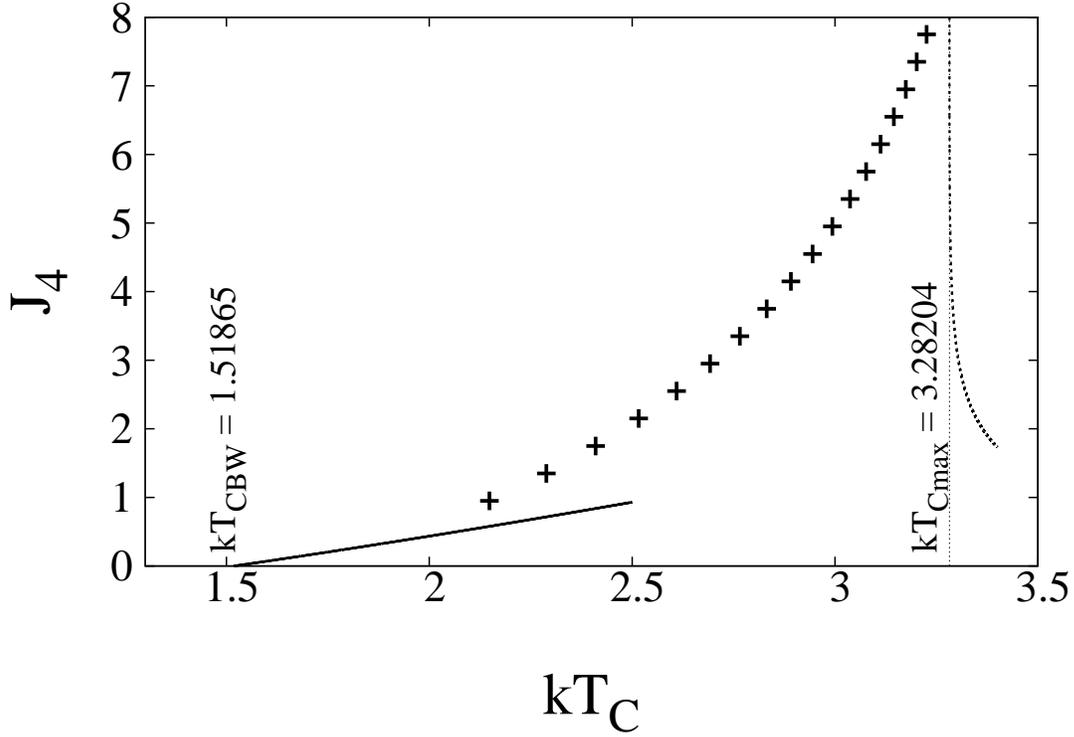}
	\caption{Relation between the critical temperature ($kT_\mathrm{C}$) and the four-spin interaction constant ($J_4$)
             in the dual model with $J_1=J_2=1$.
             The solid and dotted lines are the solution of Eqs. (\protect\ref{eq:e_cosh_tanh}) and
             (\protect\ref{eq:e_Cosh_e_sinh}), respectively,
             and the crosses indicate a possible location of the phase transition.
             $kT_\mathrm{CBW}$ is the critical temperature for the Ising model on a Brick-Wall lattice
             and $kT_\textrm{Cmax}$ is the maximum critical temperature for the ferromagnetic case of the (2 + 4) BWI model.
             }
	\label{fig:J4_T}
\end{figure}

Figure~\ref{fig:J4_T} presents a sketch of the
relation between $J_{4}$ and the critical temperature $kT_\mathrm{C}$
in a particular case of $J_{1} = J_{2} = 1$.
The solid line represents the asymptotic solution for the case $K_{4} \rightarrow 0$ obtained from \Eref{eq:e_cosh_tanh},
and the dotted line depicts the asymptotic solution of \Eref{eq:e_Cosh_e_sinh} obtained for $K_{4} \rightarrow \infty$.
The solution of \Eref{eq:e_Cosh_e_sinh}
has only the left-hand vertical asymptote at the temperature $kT_\mathrm{Cmax}$  given as the solution to
\begin{equation}
\label{eq:Cosh_e_Sinh}
  \cosh \left(\frac{2}{kT_\mathrm{Cmax}}\right)
    =  \exp\left(\frac{2}{kT_\mathrm{Cmax}}\right) \sinh \left(\frac{2}{kT_\mathrm{Cmax}}\right),
\end{equation}
which yields $kT_\mathrm{Cmax} \approx 3.28204$.
The fact that \Eref{eq:e_Cosh_e_sinh} has only the left-hand side asymptote at the critical point  defined by
$K_{4} \rightarrow \infty$
is in contradiction to the Griffith's inequalities \cite{griffiths}, which imply  that
if an additional ferromagnetic interaction is introduced into a ferromagnetic Ising model,
the critical temperature  $T_\textrm{c}$ will increase with the magnitude of this additional interaction.
Therefore, this solution has to be rejected as  nonphysical.
Our preliminary computer simulations (which will be discussed elsewhere)
suggest that the relation between  $J_{4}$ and $T_{C}$ is located along the line
marked with crosses in \Fref{fig:J4_T},
with the right vertical asymptote at the point given by \Eref{eq:Cosh_e_Sinh}.

Our considerations show that the additional four-spin interaction
in the Ising model on a Brick Wall lattice does not yield an infinite increase of the critical temperature.
Instead, the critical temperature is always limited by a condition involving two-spin interactions only.
In order to confirm this qualitative result obtained for a particular choice of $J_1$ and $J_2$,
we plan to perform detailed computer simulations of the model.

%%%%%%%%%%%%%%%%%%%%%%%%%%%%%%%%%%%%%%%%%%%%%%%%%%%%%%%%%%%%%%%%%%%%%%%
%%%%%%%%%  SECTION 5  %%%%%%%%%%%%%%
%%%%%%%%%%%%%%%%%%%%%%%%%%%%%%%%%%%%%%%%%%%%%%%%%%%%%%%%%%%%%%%%%%%%%%%

\section{Self-duality of the (2 + 4) BWI model in the presence of an external magnetic field}
As in other pairs of Ising models, the inclusion of an external magnetic field makes
the duality relationship between the NNNI  and the (2 + 4) BWI models cease to exist:
duality relations (\ref{eq:duality-UJ-BM}) -- (\ref{LLL-KKK}) were obtained only in the absence of an external magnetic field.
However, the Ising model described by the Hamiltonian (\ref{eq:H_J_J4}) has an additional internal symmetry.
If one adds to the Hamiltonian an interaction with an external magnetic field $-H \sum\limits_{i} \sigma_{i}$,
then
 the partition function will take the form
\begin{equation}
\frac{Z_{2N,2M}\left(K_{2},K_{4},h,K_{1}\right)}{2^{2MN} \left[\sinh^{2} 2K_{2}
    \left(e^{4K_{4}} \cosh^{2} 2K_{2} - 1\right)\right]^{MN/2}} =
       \operatorname{Tr} \left(V \right)^{N},
\end{equation}
where
\begin{equation}
V = e^{\tilde{K_{2}}C}e^{\tilde{K_{4}}D_{2\mathds{Z}}}e^{hE}e^{K_{1}A_{2\mathds{Z}}}
e^{\tilde{K_{2}}C}e^{\tilde{K_{4}}D_{2\mathds{Z}-1}}e^{hE}e^{K_{1}A_{2\mathds{Z}-1}}
\end{equation}
is the transfer matrix, $h = \beta H$, and
$E$ is a matrix defined as
\begin{equation}
E = \sum\limits_{k = 1}^{2M}\sigma_{k}^{x}.
\end{equation}

Let
\begin{equation}
  \mathscr{S} = \prod\limits_{i = 1}^{2M} s_{i},
\end{equation}
where $s_{i}$ are given by \Eref{eq:sj}.
This operator exchanges $\sigma_{k}^{x}$ with $\sigma_{k}^{z}$. Applying it as a similarity transformation for $V$
leads to
\begin{equation}
  V' = \mathscr{S} V \mathscr{S}^{-1},
\end{equation}
where
\begin{equation}
  V' = e^{\tilde{K_{2}}E}e^{\tilde{K_{4}}A_{2\mathds{Z}}}
          e^{hC}e^{K_{1}D_{2\mathds{Z}}}e^{\tilde{K_{2}}E}e^{\tilde{K_{4}}A_{2\mathds{Z}-1}}e^{hC}e^{K_{1}D_{2\mathds{Z}-1}}.
\end{equation}
Thus, $V'$ can be interpreted as the transfer matrix of the (2 + 4) BWI model in which the interactions have been changed
as follows,
\begin{eqnarray}
h &\rightarrow& \tilde{K_{2}},  \nonumber\\
K_{1} &\rightarrow& \tilde{K_{4}}.
\end{eqnarray}
The second difference is that while in the initial Ising model the four-spin interactions
act on the right-hand side of the bricks,
the four-spin interactions appear on their left-hand sides of the transformed model.
The self-duality relation for this  model is described by the equation
\begin{equation}
Z_{2N,2M} \left( K_{2}, K_{4}, h, K_{1} \right) =
\alpha Z_{2N,2M} \left( \tilde{K_{2}}, \tilde{K_{4}}, \tilde{h}, \tilde{K_{1}}\right),
\end{equation}
where
\begin{equation}
  \alpha = \left[\frac{\sinh^{2} 2 K_{2} \left( e^{4 K_{4}} \cosh^{2} 2 K_{2} -1 \right)}{\sinh^{2} 2 \tilde{h}
    \left(e^{4 \tilde{K_{1}}} \cosh^{2} 2 \tilde{h} - 1 \right)}\right]^{\frac{MN}{2}}
\end{equation}
and
\begin{eqnarray}
\label{eq:RSD}
 \tilde{K_{2}} &=& -\frac{1}{4} \ln \frac{\cosh 2K_{2} - e^{-2K_{4}}}{\cosh 2K_{2} + e^{-2K_{4}}}, \nonumber \\
 \tilde{K_{4}} &=& -\frac{1}{4} \ln \frac{\cosh^{2} 2K_{2} - 1}{\cosh^{2} 2K_{2} - e^{-4K_{4}}}, \nonumber\\
 \tilde{K_{1}} &=& -\frac{1}{4} \ln \frac{\cosh^{2} 2h - 1}{\cosh^{2} 2h - e^{-4K_{1}}}, \nonumber \\
 \tilde{h} &=& -\frac{1}{4} \ln \frac{\cosh 2h - e^{-2K_{1}}}{\cosh 2h + e^{-2K_{1}}}.
\end{eqnarray}

Unfortunately the model under consideration is a ferromagnetic one,
so the phase transition in an external magnetic field does not exist.
Self-duality enables one  only  to find a number of interesting relations between
correlations function for the (2 + 4) BWI model.

%%%%%%%%%%%%%%%%%%%%%%%%%%%%%%%%%%%%%%%%%%%%%%%%%%%%%%%%%%%%%%%%%%%%%%%
%%%%%%%%%  SECTION 4  %%%%%%%%%%%%%%
%%%%%%%%%%%%%%%%%%%%%%%%%%%%%%%%%%%%%%%%%%%%%%%%%%%%%%%%%%%%%%%%%%%%%%%

\section{Summary \label{sec:summary}}

The main goal of this paper was to find and investigate the dual model for the
2D Ising model with  the isotropic nearest and anisotropic next-nearest neighbor interactions  on a rectangular lattice.
The dual model  turned out to be a 2D Ising model with two-spin nearest-neighbor anisotropic
interactions and additional four-spin interactions on the Brick-Wall lattice.
The appearance of the four-spin interactions is associated with the presence of non-planar
next-nearest neighbor interactions  in the original model.
The way we constructed the dual model is quite general, does not require
the introduction of the dual lattice concept and can
be applied to other Ising models.

Investigation of the impact of the four-spin interactions on the critical temperature
of the dual model revealed that while the four-spin interactions can increase
the critical temperature of the model, this increase is restricted to a finite temperature range.
Moreover, the limits of this range are bounded by the two-spin interaction constants of the dual model.
This rather unexpected result has to be analysed further, e.g.\ by computer simulation, for example,
 using Landau's approach \cite{landau}.
Computer simulations are also necessary to investigate the exact location of the phase transition as well as its
critical properties, including the universality class.
They will also be useful in relating the results for the dual model with the properties of the original Ising model.

\ack
We would like to thank J. J\c{e}drzejewski  for his valuable comments,
especially on the important consequences of  the Griffiths inequality.
\section*{References}
%\bibliographystyle{unsrt}
%\bibliography{Publikacja_1_Eng}

\end{document}